\documentclass[11pt]{article}	
\usepackage{epsfig,amsmath,amssymb,graphics,subfigure}
\begin{document}
\title{Bounds on the Higgs Mass in Variations of Split Supersymmetry}
\author{Rakhi Mahbubani\\
\small\sl Jefferson Physical Laboratory \\
\small\sl Harvard University \\
\small\sl Cambridge, MA 02138}
\maketitle
\begin{abstract}
We investigate the limits on the higgs mass in variations of Split 
Supersymmetry, 
where the boundary value of the Higgs quartic coupling
at the SUSY breaking scale ($m_s$) is allowed to deviate from its value in the minimal model 
of Arkani-Hamed and Dimopoulos. 
We show that it is possible for $\lambda(m_s)$ to be 
negative and use vacuum stability 
to put a lower bound on this coupling, 
and hence on the mass of the physical higgs.
We also use the requirement of perturbativity of all couplings up to the cutoff  
to determine an upper limit for the higgs mass in models which are further modified by additional matter content.  
For $m_s\geq 10^6$ GeV we find $110$ GeV $\lesssim m_h\lesssim 280$ GeV if the new matter is not coupled to any Standard Model field; and $110$ GeV $\lesssim m_h\lesssim 400$ GeV  if it has Yukawa couplings to the higgs. 
\end{abstract}

\section{Introduction}

The quest for a natural way to break electroweak symmetry has long 
been the central motivation 
for constructing theories for physics beyond the Standard Model (SM) at the TeV scale. However, the same logic applied to the even more severe fine-tuning problem associated with the cosmological constant would have predicted new physics near $10^{-3}$ eV, which we have no evidence for. This suggests
the possibility that our notions of naturalness are misleading, and that other fine-tuning mechanisms may be at work in nature.  

Eliminating the use of naturalness as a guiding 
principle for model-building allows for some drastic changes to particle physics lore.  Arkani-Hamed and Dimopoulos have recently argued 
for a theory with ``Split'' supersymmetry \cite{Arkani-Hamed:2004fb} (also 
\cite{Giudice:2004tc}). In this model the ``structure'' \cite{Weinberg:1987dv} and ``atomic''  \cite{Agrawal:1997gf} principles were used to explain the smallness of UV sensitive parameters; namely the cosmological constant and the higgs vacuum expectation value.  Chiral symmetries keep all fermions light and a single fine tuning does the same for one higgs scalar, while all other scalars are at the high SUSY breaking scale.  This framework,  while preserving the successes of the MSSM such as gauge coupling unification \cite{Dimopoulos:1981zb,Marciano:1981un}, 
also salvages some of its difficulties, giving a simple explanation for the absence of FCNCs and CP violation; 
the non-discovery of superpartners, and a light higgs (see also \cite{Wells:2003tf}), while simultaneously solving a variety of cosmological difficulties associated with the gravitino and moduli. 

An important quantitative prediction of the minimal model is the mass of the higgs, which has been computed to lie between 120 and 170 GeV \cite{Arvanitaki:2004eu}. As pointed out in \cite{Arkani-Hamed:2004fb}, however, this prediction is sensitive 
to physics above $m_s$ and the presence of new matter beneath $m_s$. In this paper we explore the bounds on the Higgs mass in these 
more general versions of Split SUSY which continue to conform to the essential philosophy of the model, in order to provide a falsifiable test of theories that are built on these principles. 
We analyse these models at one loop and examine the limits on the boundary value of the higgs quartic coupling at the SUSY breaking scale.  
In Section \ref{sec:minmass} we show that there exist mechanisms by which this can be 
made negative.  How negative is determined by requiring stability of the SM vacuum. 
 Next, in analogy with the triviality bound in the SM, we use the requirement of 
perturbativity to the cutoff to put an upper limit on this boundary value.  
We take a totally agnostic viewpoint, assuming that there is some unknown physics in the region 
between $m_s$ and the unification scale, $m_{GUT}$, that effectively decouples the higgs quartic 
from the electroweak gauge couplings.  This could include new D-terms or F-terms such as in the NMSSM.  
We then see how far this bound can be pushed in the large $\tan{\beta}$ limit by: 
\begin{itemize}
\item{varying $m_s$ and adding N SU(5) $(\bf{5}+\bf{\overline{5}})$s at the weak scale, maximizing N at each $m_s$ in order to maintain perturbative gauge coupling unification}
\item{adding yukawa couplings between the higgs and new fermions that come in complete multiplets of SU(5)}
\end{itemize}
We consider each of these in turn in Section \ref{sec:maxmass}, RGE evolving the couplings down to the weak scale, where we can calculate the physical higgs mass.  This gives us a firm prediction of this class of models, based on a minimal number of reasonable assumptions.  We conclude in Section \ref{sec:conclusion}, with some discussion of possible interesting extensions of this work.

\section{Lower limit on the higgs mass}\label{sec:minmass}

We begin by discussing how we can {\it decrease} the higgs quartic coupling at $m_s$.  Suppose that in the theory above this scale, we have an additional gauge singlet scalar field $N$ that picks up a mass term $m_s^2 |n|^2$ from SUSY breaking.  Like the higgses, it has an $R$-charge of zero so that the superpotential term 
$N H_u H_d$ is forbidden.  However, the following $A$-term is permitted 
\begin{equation}\label{equ:softterms}
V_{soft}\supset m_snh_uh_d 
\end{equation}
Integrating out $n$ induces a term of the form $-(h_u h_d)^2$, and 
our effective theory beneath $m_s$ now contains a negative contribution to the higgs quartic coupling of
\begin{equation}\label{equ:negativequartic}
 -\sin^2{2\beta}(h^\dag h)^2.
\end{equation}
This can exceed the usual gauge D-term contribution proportional to $\cos^2{2\beta}$, 
giving rise to the possibility that the SM vacuum state is not the true vacuum of the theory and allowing for the eventual decay of 
\begin{figure}[thbp]
\begin{center}
\epsfig{file=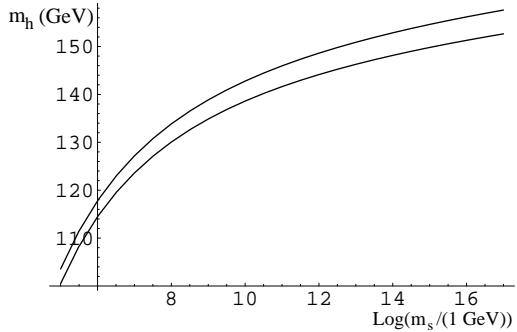,width=7cm}
\caption{Minimum possible higgs mass given the need for vacuum stability.  The upper line is for $\tan{\beta}=50$ and the lower line is for $\tan{\beta}=1$.}
\label{fig:minmass}
\end{center}
\end{figure}
our vacuum to the true one by bubble nucleation.  We use the methods in \cite{Coleman:1995} to calculate the decay rate per unit volume by approximating this to a pure $-\phi^4$ theory (see e.g \cite{Isidori:2001bm}):
\begin{equation}\label{equ:decayrate}
\frac{\Gamma}{V}\simeq \int \frac{dR}{R} \left(\frac{1}{R}\right)^4e^{-\frac{16\pi^2}{3|\lambda(1/R)|}}
\end{equation}
where $R$ is the size of the bubble by which this process takes place. 
In practice this integral is just dominated by the scale at which the integrand is maximized. As we will see, $|\lambda(\mu)|$ will 
turn out to be largest for $\mu \sim m_S$, so the rate is dominated by $R \sim m_S^{-1}$.   
We musy have $\Gamma/V \lesssim (\text{Hubble})^4$ for this decay not to have occurred already, 
and solving this equation allows us to bound $\lambda(m_s)$:
\begin{equation}\label{equ:lowlambdabound}
\lambda(m_s)\geq -\frac{4\pi^2}{3\ln{(\frac{m_s}{\text{Hubble}})}}
\end{equation}
Saturating this bound we find the higgs mass shown in Figure \ref{fig:minmass} for two different values of $\tan{\beta}$.  Note the consistency of this lower limit on the higgs mass for $m_s=10^6$ GeV with the LEP-II bound.  We do not wish to lower $m_s$ any further since this will bring back the problems associated with the MSSM that we were trying to alleviate.  

\section{Upper limit on the higgs mass}\label{sec:maxmass}

Adding matter at the weak scale does not disrupt one-loop gauge coupling unification as long as this matter can be grouped into complete multiplets of SU(5).  It does, however, increase the value 
\begin{figure}[htbp]
\begin{center}
\epsfig{file=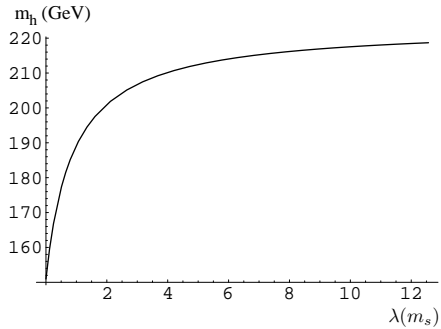,width=6cm}
\caption{Higgs mass as a function of boundary value for quartic coupling $\lambda$. \label{fig:higgsmassvslambda}}
\end{center}
\end{figure}
of the couplings at $m_{GUT}$ since it contributes (the same) positive quantity to each RGE.  Hence we need to ensure that we do not add so many particles that the couplings become non-perturbative before unification takes place.  
\begin{figure}[t]
\begin{center}
\epsfig{file=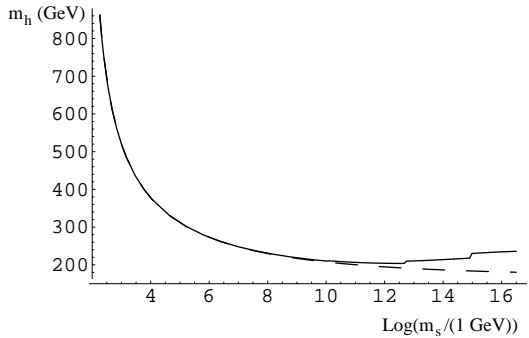,width=7cm}
\caption{Maximum higgs mass as a function of SUSY breaking scale adding N $(\bf{5}+\bf{\overline{5}})$s of SU(5).  This can be compared with the result without any extra GUT multiplets (dashed line).  The higgs quartic is taken close to its perturbative limit at the cutoff. \label{fig:higgsmassnplusnbar}}
\end{center}
\end{figure}
This limits N$\leq$6 for SUSY breaking scale around $10^9$ GeV,  for example.  We use these values to show in Figure \ref{fig:higgsmassvslambda} how the low energy physical higgs mass changes with $\lambda(m_s)$. Note that there is very little gain in mass for $\lambda > 4$ at high energies.

Next we take the higgs quartic close to its perturbative limit at the cutoff and vary $m_s$, maximizing N at each scale to find an upper bound on the higgs mass.  We expect the bound to increase substantially with N; however as can be seen in Figure \ref{fig:higgsmassnplusnbar} this is not exactly the case; at least for low cutoff there is no significant difference between the higgs mass in the theory with and without extra $(\bf{5}+\bf{\overline{5}})$s.\footnote{Without extra multiplets the results are almost identical to the SM triviality limit.  The discontinuities in the higgs mass for the high cutoff region correspond to energies where N increases by one.}  This seems rather counterintuitive since increasing all gauge couplings increases the boundary values of the higgs-gaugino yukawa couplings ($\kappa$s) at the SUSY breaking scale, which in turn should feed into the higgs mass.  
\begin{figure}[htbp]
\centering \subfigure[$m_s$=$10^{8}$]{
\label{fig:lowcutoff} \centering
\includegraphics[width=.45\textwidth]{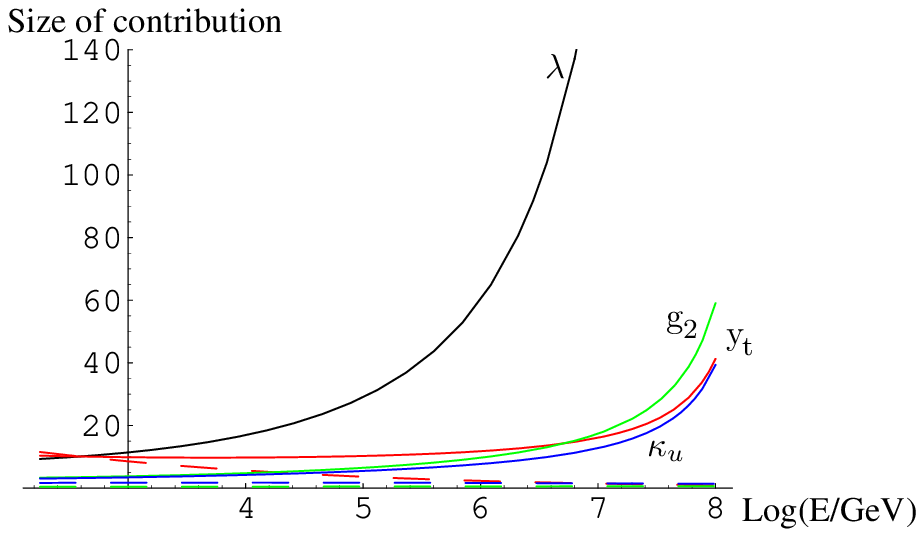}}
\subfigure[$m_s$=$10^{16}$]{\label{fig:highcutoff}
\centering
\includegraphics[width=.45\textwidth]{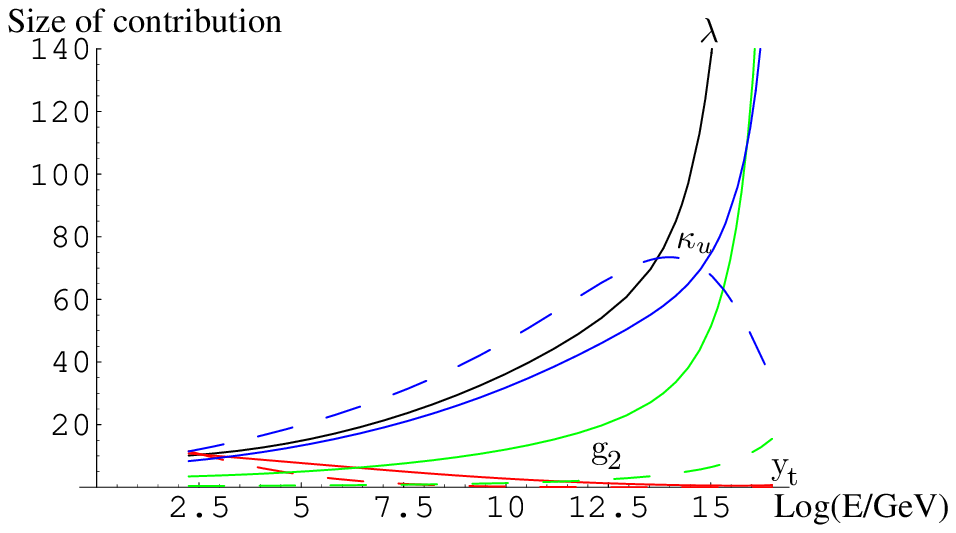}}
\caption{Absolute value of dominant contributions to the one-loop running of the higgs quartic for large $\lambda(m_s)$.  The $\lambda$ contribution is black, $y_t$ is red, $\kappa_u$ is blue and $g_2$ is green; dashed lines correspond to $(\text{coupling})^4$ terms.} 
\label{fig:contrib}
\end{figure}
The reality of the story for $\lambda$ is rather more complicated, however, and intimately involves three other couplings, $y_t$, $\kappa_u$ and $g_2$, in the terms $\lambda (\text{coupling})^2$ as well as $\text{(coupling)}^4$.  Figure \ref{fig:contrib} contains a graph of each of these contributions to the $\lambda$ RGE.  Notice that for low cutoff, the running of $\lambda$ is dominated by itself and, since we have decoupled its boundary condition from the electroweak gauge couplings, it relies on none of the quantities that change on adding $(\bf{5}+\bf{\overline{5}})$s.  

Even if $\lambda$ was not the dominant coupling, increasing $\kappa_u$ would actually decrease the higgs mass since it contributes via the positive $\lambda\kappa_u^2$ term, partly undoing the effects of increasing $g_2$ and decreasing $y_t$.  For high cutoff on the other hand, the quartic runs down enough so that not only do the weak gauge coupling and gaugino yukawa start playing a much bigger part in determining its running (although $y_t$ still does not), but the $-\kappa_u^4$ term actually becomes the dominant one.  Increasing N therefore increases the higgs mass through larger $\kappa_u$ as well as $g_2$.  Due to this property of the RGEs in this theory our method of adding $(\bf{5}+\bf{\overline{5}})$s will not increase the higgs mass much beyond 250 GeV. 
\begin{figure}[htbp]
\centering\subfigure[MSSM boundary condition]{\label{fig:NMSSMbc}
\label{fig:MSSM} \centering
\includegraphics[width=.45\textwidth]{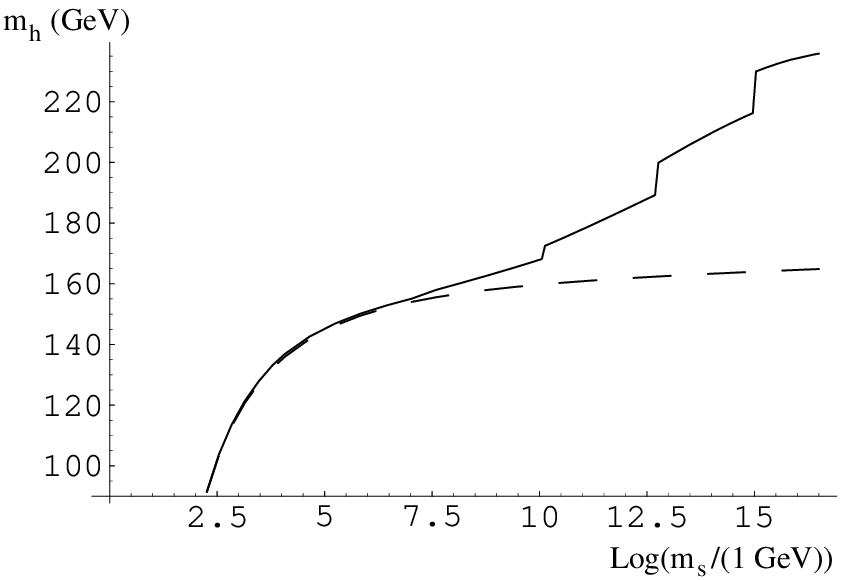}}
\subfigure[New Fat Higgs boundary condition]{\label{fig:newfatbc}
\centering
\includegraphics[width=.45\textwidth]{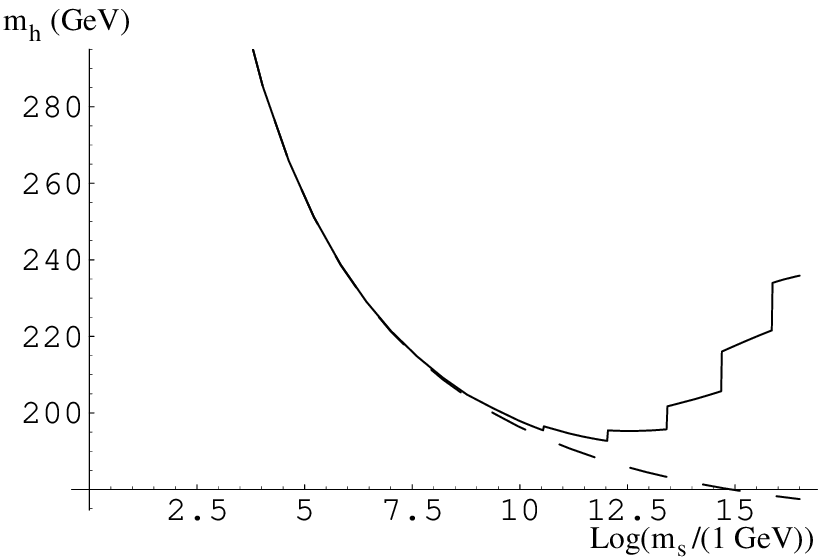}}
\caption{Maximum higgs mass as a function of SUSY breaking scale with different boundary condition for the higgs quartic at $m_s$.} \label{fig:higgsmassnplusnbarlowbc}
\end{figure}
As a check we can see in Figure \ref{fig:higgsmassnplusnbarlowbc} similar results using the boundary conditions for the MSSM and the New Fat Higgs \cite{Chang:2004db} respectively.  Since these are much smaller than the perturbative limit used in the previous example, the results with and without new GUT multiplets start to differ at a lower energy, confirming that $\lambda$ now does not need to run down as much before the other couplings start becoming important.\footnote{In UV completing Split SUSY with the New Fat Higgs we took $m_s = \Lambda = M_X = M_{\tilde{X}}$, although in reality the latter two need to be separated slightly in order for us to legitimately use the weak limit bound described in the paper.  Recall that this model already contains 4 $(\bf{5}+\bf{\overline{5}})$s of its own above the confinement scale - we need to take these into account in our perturbative unification constraint.}  

Returning to the effect of the top yukawa, we saw that this was negligible for all cutoffs since its value at low energies is fixed, and so it never becomes comparable in size to the other terms in the $\lambda$ RGE.  This observation inspires an alternative approach in which N vector-like GUT multiplets of quarks and leptons ($(\bf{5}+\bf{\overline{5}}+\bf{10}+\bf{\overline{10}})$s) are added and coupled to the higgs in the usual fashion.  These have vector-like masses, and the new top-like yukawa coupling, $y_t'$, plays exactly the same role as $y_t$, except that it is not fixed at low energies and therefore can be more instrumental in determining the higgs mass.

\begin{figure}[h]
\begin{center}
\epsfig{file=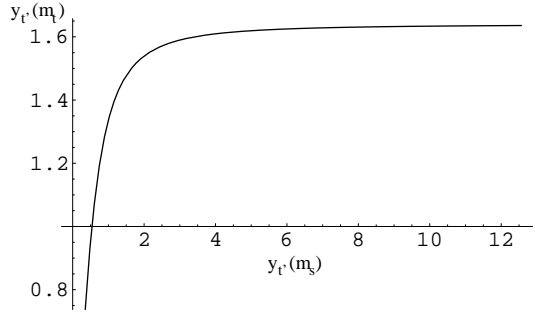,width=7cm}
\caption{Low energy value of the new top' yukawa coupling as a function of its value at the SUSY breaking scale. \label{fig:ytvsyt}}
\end{center}
\end{figure}
\begin{figure}[h]
\begin{center}
\epsfig{file=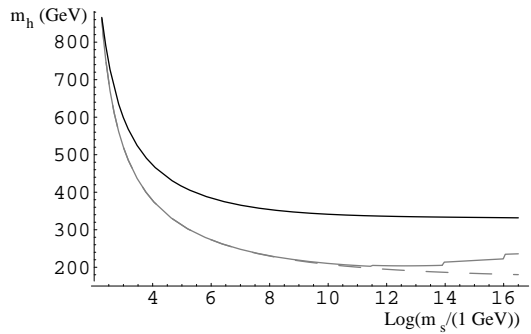,width=7cm}
\caption{Maximum higgs mass as a function of SUSY breaking scale for Split SUSY with N additional $(\bf{5}+\bf{\overline{5}}+\bf{10}+\bf{\overline{10}})$s of SU(5) (solid black line) for $y_{t}'(m_s)=4$.  Our previous results with (solid grey line) and without (dashed gray line) additional $(\bf{5}+\bf{\overline{5}})$s are also shown for comparison. \label{fig:maxhiggsmassxtend}}
\end{center}
\end{figure}

First we examine how the low energy value of this additional yukawa changes with boundary condition at the cutoff (see Figure \ref{fig:ytvsyt}).  As with $\lambda$, it is relatively insensitive to its boundary condition for $y_{t}'(m_s)\geq$2.  For such a large yukawa the higgs mass is also independent of the boundary value of $\lambda$, suggesting that its RGE is now controlled by the new yukawa as intended.  Now we can analyse the higgs mass with changing cutoff in a similar manner to the ($\bf{5}+\bf{\overline{5}}$) case from earlier, maximizing N at each energy and comparing with our previous results in Figure \ref{fig:maxhiggsmassxtend}.  If we examine the different contributions to the running of $\lambda$ as before, we see that the larger $y_{t}'$ indeed plays a leading role along the entire cutoff spectrum, giving rise to an overall increase in higgs mass with N, to a maximum of about 400 GeV.

It is possible that, as it stands, this modification with a dominant top-type yukawa coupling will give rise to undesirably large oblique parameters, especially a large positive contribution to T.  This issue can be resolved in two ways, neither
 of which significantly affect our results.  Firstly we could increase the masses of these new fermions, which suppresses the higher-dimensional operators contributing to precision electroweak measurements like the T parameter.  Alternatively, we could impose an approximate custodial $SU(2)$ symmetry in the new matter sector, fixing the same boundary conditions on the bottom-type yukawa as the top-type.  Either way, these models can be made consistent with current experimental data.

\section{Conclusion}\label{sec:conclusion}
We see that it is possible to give limits on the physical higgs mass of $110\text{ GeV}\leq m_h\leq 250\text{ GeV}$ in arbitrary extensions of Split SUSY, including additional GUT matter multiplets/GUT singlets that are otherwise decoupled from the SM.  Adding new yukawa couplings pushes the upper bound to around $400$ GeV.  It is interesting that the {\it lower} limit on the higgs mass has already 
been excluded by LEPII.  However, we see that the possible hint for a 115 GeV higgs can be accomodated in Split SUSY models with $m_s 
\sim 10^6$ GeV and a slightly negative Higgs quartic coupling near this scale.  

Further analysis is required to determine how robust these predictions are to adding two-loop contributions to the running (see \cite{Falck:1985aa,Machacek:1984zw} for RGEs). 
 From the corresponding SM results \cite{Altarelli:1994rb}, we expect that this will decrease our limits by of the order of $10$ GeV.  It would also be interesting to see how NMSSM-like boundary conditions for the higgs quartic, or even different UV completions of the NMSSM, such as \cite{Harnik:2003rs}, affect these bounds.

\section*{Acknowledgements}
Many thanks to Nima Arkani-Hamed for suggesting the possibility of a negative quartic and also for numerous invaluable discussions on the vacuum stability bound among other interesting things.

\end{document}